\documentclass[twocolumn]{aastex631}
\hypersetup{linkcolor=red,citecolor=blue,filecolor=cyan,urlcolor=magenta}

\newcommand{\kms}{$\rm{km\,s^{-1}}$}
\newcommand{\vlsr}{$v_{\rm LSR}$}
\newcommand{\absvlsr}{$|v_{\rm LSR}|$}
\newcommand{\cmsq}{$\rm{cm^{-2}}$}
\newcommand{\msun}{$\rm{M_\odot}$}

\newcommand{\hi}{\ion{H}{1}}

\defcitealias{Bordoloi2017}{B17} 

\shorttitle{Cool HVCs in the Fermi Bubbles}
\shortauthors{Bordoloi et al.}

\begin{document}

\title{A New High-latitude \hi\ Cloud Complex Entrained in the Northern Fermi Bubble}

\correspondingauthor{Rongmon Bordoloi}
\email{rbordol@ncsu.edu}\tabletypesize{\footnotesize}

\author[0000-0002-3120-7173]{Rongmon Bordoloi}
\affiliation{Department of Physics, North Carolina State University, Raleigh, 27695, North Carolina, USA}

\author[0000-0003-0724-4115]{Andrew J. Fox}
\affiliation{AURA for ESA, Space Telescope Science Institute, 3700 San Martin Drive, Baltimore, MD 21218}

\author[0000-0002-6050-2008]{Felix J. Lockman}
\affiliation{Green Bank Observatory, National radio Astronomy Observatory, Green Bank, WV 24944, USA}

\begin{abstract}

We report the discovery of eleven high-velocity \hi\ clouds at Galactic latitudes of 25--30 degrees, likely embedded in the Milky Way’s nuclear wind. The clouds are detected with deep Green Bank Telescope 21\,cm observations of a $3.2^\circ\times6.2^\circ$ field around QSO 1H1613-097, located behind the northern Fermi Bubble. Our measurements reach $3\sigma$ limits on $N_{\rm HI}$ as low as $3.1 \times 10^{17}$ cm$^{-2}$, more than twice as sensitive as previous \hi\ studies of the Bubbles. The clouds span $-180 \leq v_{\rm LSR} \leq -90$ \kms\ and are the highest-latitude 21 cm HVCs detected inside the Bubbles. Eight clouds are spatially resolved, showing coherent structures with sizes of 4--28 pc, peak column densities of log($N_{\rm HI}$/cm$^{2}$)=17.9--18.7, and \hi\ masses up to 1470\,\msun. Several exhibit internal velocity gradients. Their presence at such high latitudes is surprising, given the short expected survival times for clouds expelled from the Galactic Center. These objects may be fragments of a larger cloud disrupted by interaction with the surrounding hot gas.

\end{abstract}


\section{Introduction} \label{sec:intro}

The Milky Way (MW) drives a nuclear wind from the Galactic Center (GC) 
into the halo, providing an important channel by which mass and energy 
are circulated around the Galaxy. Understanding and characterizing 
the nuclear wind is an important part of the broader effort to understand 
the Galactic ecosystem and baryon cycle across cosmic time \citep{Tumlinson2017,Chen2024,Bordoloi2024}. 
Hot outflowing gas is seen in the form of the Fermi Bubbles \citep{Su2010, Sarkar2024}, 
two giant plasma lobes extending more than 10 kpc into both Galactic hemispheres, 
as well as their larger X-ray counterparts, the eROSITA bubbles \citep{Predehl2020}.
The bubbles emit radiation from gamma rays \citep{Ackermann2014}, through X-rays and 
the mid-IR \citep{BH03, Predehl2020}, to microwaves and radio waves
\citep{Finkbeiner2004, Carretti2013}.

Embedded within the wind is a population of clouds 
seen in the form of high-velocity clouds (HVCs) 
in \hi\ 21 cm emission \citep{MG13, dT18, Lockman2020, Gerrard2024},
CO emission \citep{DiTeodoro2020, Noon2023, Heyer2025}, molecular $H_2$ absorption \citep{Cashman2021}, 
and UV metal-line absorption 
\citep{Keeney2006, Fox2015, Savage2017, Karim2018,Cashman2023}
in sightlines passing through the Fermi Bubbles. 
These Fermi Bubble HVCs are thus multi-phase gas clouds ($T\!\sim\!10^{2-5.5}$ K) embedded in the nuclear wind.
The UV and radio studies have revealed many properties of the Fermi Bubble HVCs, 
including their kinematics \citep[][hereafter \citetalias{Bordoloi2017}]{Bordoloi2017}, 
\hi\ masses \citep{dT18}, 
covering fraction \citep{Ashley2020}, and chemical composition \citep{Ashley2022}.
The kinematics of the UV absorbers \citepalias{Bordoloi2017} and the \hi\ emitters
\citep{dT18, Lockman2020} are consistent with a biconical wind launched 
from the Galactic Center several Myr ago.

Until now, the \hi\ HVCs in the Fermi Bubbles have been found at low latitude
\citep[$|b|<10^\circ$;][]{dT18}. In this Letter, we report the discovery of a 
complex of \hi\ clouds at much higher latitude, in the range $b=25-30^\circ$.
We discuss the implications of the existence of neutral clouds in the Fermi Bubbles
at such distance from the Galactic plane.

\begin{figure*}
\centering
\includegraphics[width=2\columnwidth]{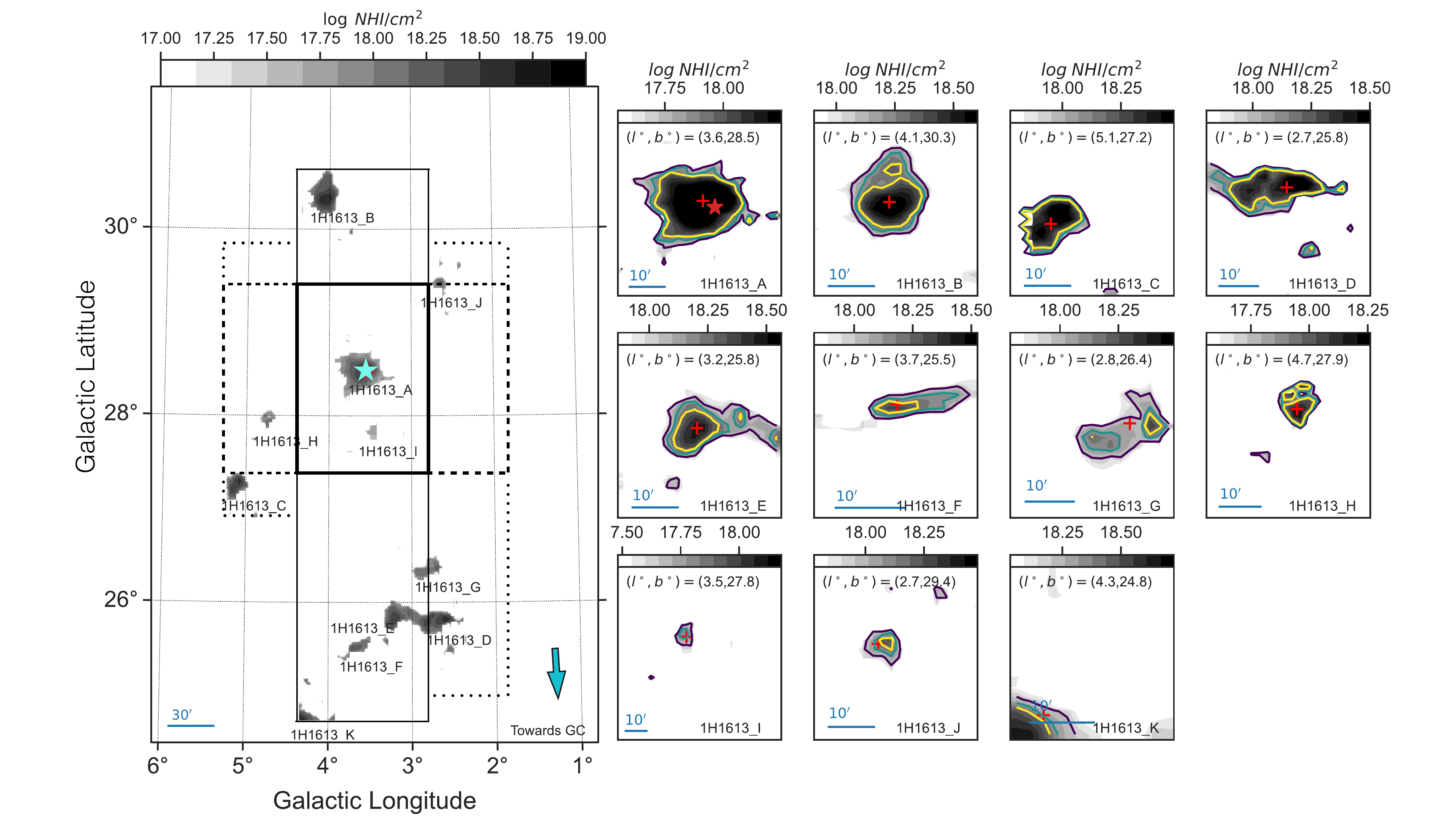}
\caption{GBT \hi\ maps of the region surrounding the background QSO 1H1613-097 (cyan star). The {\bf left panel} displays the full $\approx3.2\times6.2^\circ$ field covered by our observations. Thick black lines, thin black lines, dotted, and dashed lines outline regions where the median rms brightness temperature noise in a 1 \kms\ channel is 10, 15, 20, and 30 mK, respectively. The {\bf right panel} presents insets of the eleven individual clouds, labeled A–K, with overlaid contours of $N_{\rm HI}$. All maps show \hi\ emission integrated over the velocity range $-170 \leq v_{\rm LSR} \leq -90$ \kms. Red crosses indicate the $T_{\rm B}$-weighted spatial coordinates of each cloud, while the red star in the panel containing cloud A marks the location of the background quasar. The spatial proximity and similar kinematics of the seven clouds suggest a physical association. The color contours represent the 3, 4, and 5$\sigma$ column density isophotes in each frame.}
\label{fig:GBT_map}
\end{figure*}

\section{Observations and Data Reduction} \label{sec:obs}

We performed 21\,cm \hi\ measurements  of an area $3.2^\circ\times6.2^\circ$ around the 
UV-bright quasar QSO 1H1613-097 using the Robert C. Byrd Green Bank Telescope (GBT) of the Green Bank Observatory under program numbers GBT 16B-422, GBT 17B-015 and GBT 20A-253.

The QSO is located at galactic coordinates $l,b =3.6^{\circ},+28.5^{\circ}$, 
behind the center of the northern Fermi Bubble.
The field around this quasar was chosen because HST/COS spectroscopy of 
QSO 1H1613-097 revealed complex multi-phase high-velocity gas in 
absorption, tracing diffuse ionized plasma at temperatures of $10^{4-5}$K
\citepalias{Bordoloi2017}. Earlier 21\,cm GBT observations made towards the QSO 1H1613-097 
detected an HVC coincident with the velocity of the UV metal absorption lines at 
\vlsr=$-$172\,\kms\ \citepalias{Bordoloi2017}. 

The GBT observations were made ``on-the-fly'' repeatedly covering an area until the desired sensitivity was reached.
The total observing time was $\approx 114$ hours.
The data were taken by frequency-switching using the VEGAS spectrometer \citep{Prestage2015}. 
The spectra were calibrated and corrected for stray radiation as described in \cite{Boothroyd2011}, were smoothed to an effective velocity resolution of 0.94 \kms, and assembled into a cube that  covered $-$478 $\leq$ \vlsr $\leq +300 $ \kms\ using the \texttt{gbtgridder} software.
An instrumental baseline was removed from each pixel using a fifth-degree polynomial.
At the frequency of the 21cm line the GBT beam has a FWHM of $9.1\arcmin$;
the effective angular resolution of the final data cube is $10.0\arcmin$.

The detailed coverage of the observations is shown by the lines in the left panel of Figure~\ref{fig:GBT_map}.
The observing program adopted a ``wedding-cake'' strategy, creating regions of varying sensitivity across the mapped area.
Noise levels are expressed as brightness temperature in a 1 \kms\ velocity channel.
The main region, covering $2.73^\circ \leq \ell \leq 4.35^\circ$ and $24.8^\circ \leq b \leq 30.7^\circ$, was mapped to an rms noise level of 15 mK.
Within this area, a zone centered on the QSO—bounded by $2.73^\circ \leq \ell \leq 4.35^\circ$ and $27.7^\circ \leq b \leq 29.2^\circ$ (outlined by the thick black line in Figure~\ref{fig:GBT_map})—was mapped to a higher sensitivity with an rms noise of 10 mK.
Two additional regions adjacent to the main area, delineated by dashed and dotted lines, were mapped to rms noise levels of 20 mK and 30 mK, respectively.
These sensitivities correspond to 3$\sigma$ \hi\ detection limits for a 30 \kms\ FWHM line of $N_{\rm HI} = 3.1\times10^{17}$\,cm$^{-2}$ and $4.6\times10^{17}$\,cm$^{-2}$ in the most sensitive regions, and $6.1\times10^{17}$\,cm$^{-2}$ and $9.2\times10^{17}$\,cm$^{-2}$ in the outer regions.

These measurements are more than twice as sensitive as previous studies of \hi\ emission in the Fermi Bubbles, which had typical rms noise levels of 23--40 mK \citep{dT18, Lockman2020}.

\begin{figure*}
\centering
\includegraphics[width=1.5\columnwidth]{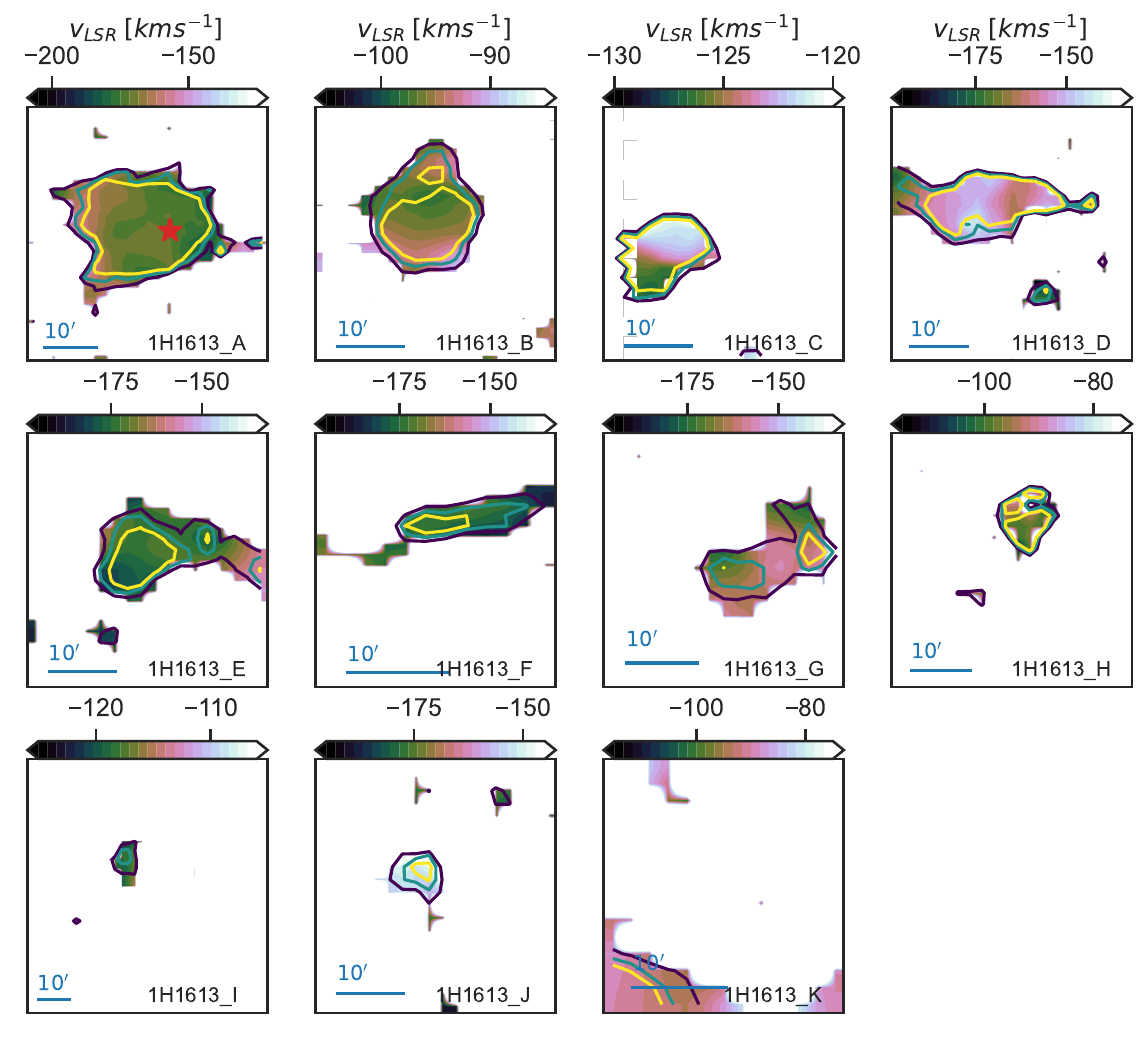}
\caption{GBT first-moment maps of the HVCs detected in Figure \ref{fig:GBT_map}, showing the mean $v_{\rm LSR}$ per spaxel for each cloud. The colored contours represent the 3, 4, and 5$\sigma$ column-density isophotes for each cloud. The red star in the panel containing cloud A marks the location of the background quasar.}
\label{fig:GBT_map_kinematics}
\end{figure*}

\begin{figure*}
\centering
\includegraphics[width=1.95\columnwidth]{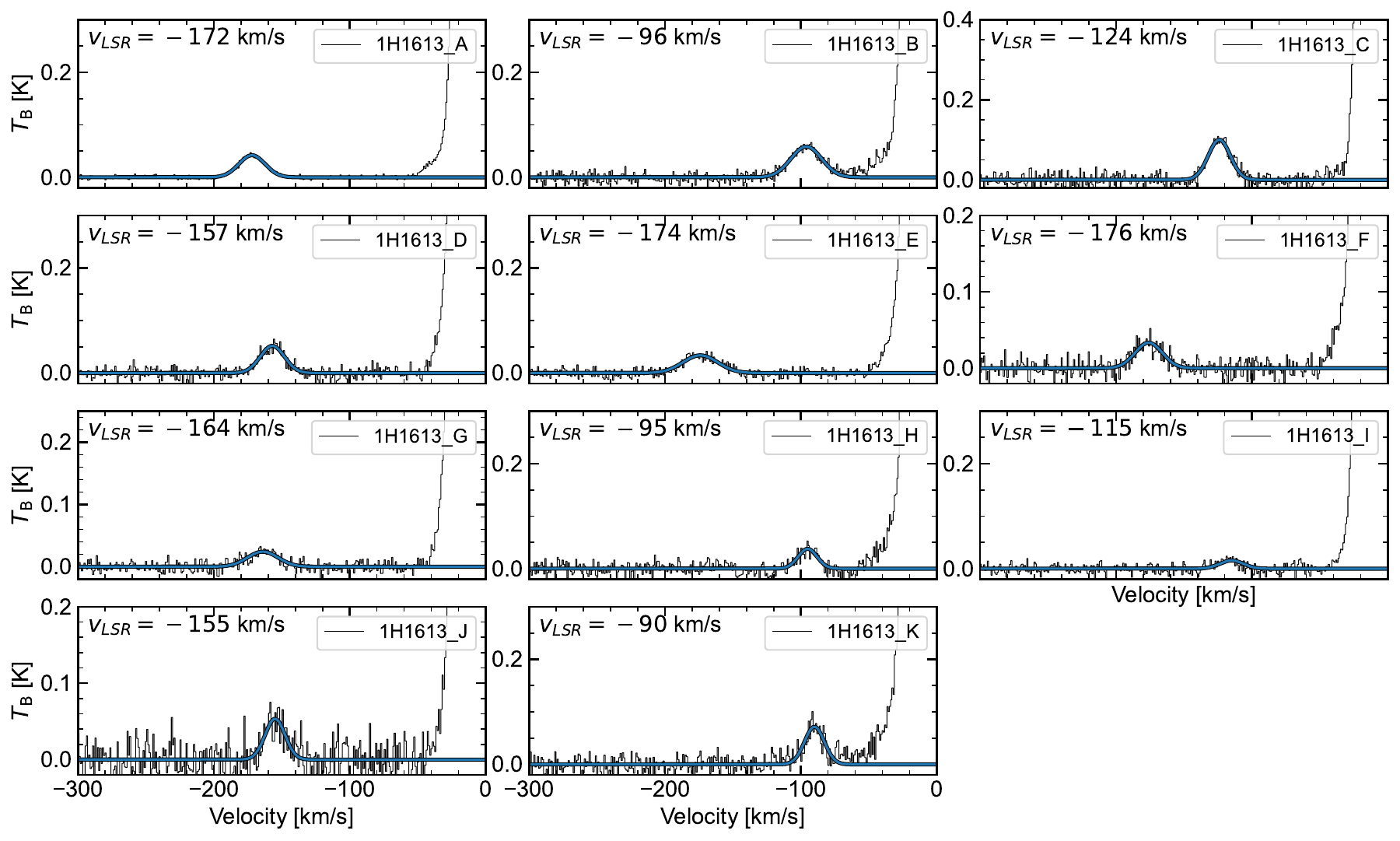}
\caption{Brightness temperature–weighted 1D spectra for each cloud shown in Figure~\ref{fig:GBT_map_kinematics}. Gaussian fits are overplotted as solid blue lines. The sharp rise in $T_B$ at $v_{\rm LSR} \gtrsim -50$~\kms\ reflects contribution from the Milky Way's interstellar medium. The mean velocity of each cloud is labeled in the corresponding panel.} 
\label{fig:GBT_1d_spec}
\end{figure*}

\begin{figure}
\centering
\includegraphics[width=0.95\columnwidth]{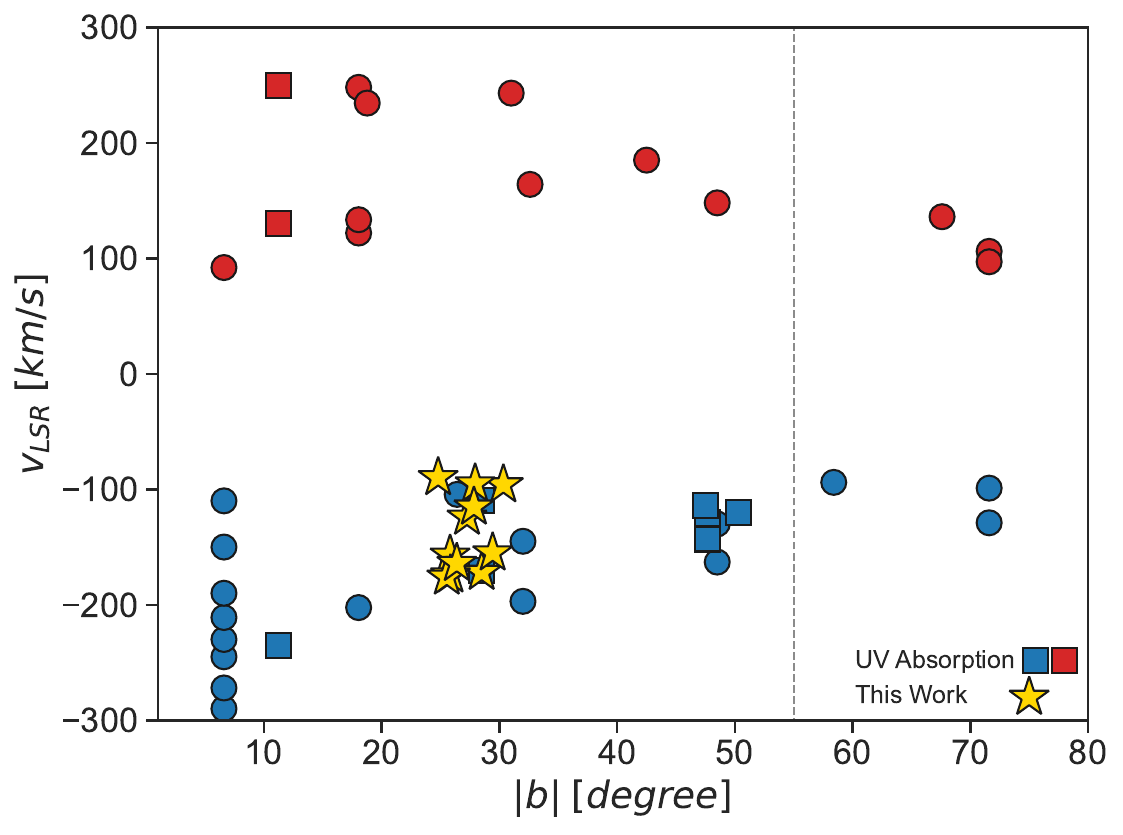}
\caption{Velocity profile of Fermi Bubble HVCs seen in UV absorption. LSR velocity 
is plotted against absolute latitude. The blue and red points show the FB HVCs from the 
sample of \citet{Ashley2020}, with circles showing northern directions and squares showing
southern directions. The vertical dashed line marks the height of the Fermi Bubbles as seen in gamma-ray emission. The yellow stars indicate the eleven new clouds presented in this work. By definition, HVCs must have $|v_{\rm LSR}| > 90$~\kms. A trend of decreasing $|v_{\rm LSR}|$ with increasing $|b|$ is evident at both positive and negative velocities, and at both northern and southern latitudes.}
\label{fig:lsr_profile}
\end{figure}

\section{Cloud Identification and Measurements}\label{sec:cloudid}

We visually inspected the baseline-subtracted data cube and identified eleven individual HVCs at \absvlsr\ $>$90 \kms, labeled Clouds A through K.
To quantify the detection significance of these clouds, we computed the noise statistics for each individual spaxel.
Following \cite{dT18}, we calculated the noise level as $\sigma_{\rm rms} = 1.4826\times \epsilon$, where $\epsilon$ is the median absolute deviation (MAD) from the median brightness spectrum of each spaxel.
We considered a cloud to be robustly detected if at least five adjacent spaxels exhibited brightness temperatures $T_{\rm B}>3\sigma_{\rm rms}$, and if the cloud spanned at least 3 \kms\ in spectral width (i.e., three spectral channels).
Thumbnail images of these eleven detected HVCs are shown in the right panel of Figure~\ref{fig:GBT_map}.  All detected \hi\ emission clouds have \vlsr $ < -90$ \kms\ even though the spectra were equally sensitive to positive velocity emission.  The implications of this are discussed in section \ref{sec:results}.

We extracted the $T_{\rm B}$ spectrum of each HVC to determine the velocity range over which it is detected ($v_{\rm min}$ to $v_{\rm max}$).
This range was used to create a zeroth-moment $T_{\rm B}$ map, which served as a weight map to extract $T_{\rm B}$-weighted spectra. 
Each weighted spectrum was fit with a Gaussian profile to determine the mean velocity (\vlsr) and velocity dispersion ($\Delta v$); these values are reported in Table~\ref{:tab:cloud_properties}. We also estimated the velocity spread within each cloud by calculating the $\Delta v_{90}$ statistic, defined as the range between the 5th and 95th percentiles of the velocity distribution for all statistically significant spaxels within the cloud.
In addition, we computed the first moment of $T_{\rm B}$ in each spaxel to map the mean velocity distribution of the HVCs, producing the velocity maps shown in Figure~\ref{fig:GBT_map_kinematics}.
Cloud positions were defined as the $T_{\rm B}$-weighted galactic latitudes and longitudes (i.e., the center of mass of each cloud). These coordinates are adopted throughout this work and are reported in Table~\ref{:tab:cloud_properties}.

We identified the spaxel with the highest $T_{\rm B}$ in each cloud and extracted the corresponding spectrum. A Gaussian profile was fit to determine the peak $T_{\rm B}$ ($T_{\rm B,max}$) and the maximum \hi\ column density ($\log N_{\rm HI,max}$) in this spaxel, assuming the gas is optically thin. The \hi\ column density is defined as 

\begin{equation}
N_{\rm HI} =1.82\times10^{18}\times \int_{v_{\rm min}}^{v_{\rm max}} \frac{T_{\rm B} (v)}{\rm K} \frac{dv}{\rm km\;s^{-1}} [{\rm cm}^{-2}].  
\end{equation}

Cloud sizes were derived following the framework presented in \cite{dT18}. 
We make the simplifying assumption of circular cloud geometry and derive the cloud 
sizes by $r_{cl}=D \times \tan{\sqrt{A/\pi}}$. Here, $D$=7.5 kpc is the distance 
to the clouds from the observer, estimated from kinematic modeling of the 1H1613+097 
HVC cloud \citepalias{Bordoloi2017}. $A$ is the angular area covered by all detected pixels 
in the zero-moment maps after accounting for the GBT beam size. The final \hi\ mass 
associated with each cloud is computed under the assumption that the gas is optically 
thin, $M_{HI}=\mu m_{p} N_{HI} A$, where $\mu m_{p}$ is the mean atomic weight of 
hydrogen. All measurements are reported in Table \ref{:tab:cloud_properties}.

\section{Results and Discussion}\label{sec:results}

We detect eleven distinct 21 cm HVCs in the region surrounding the QSO 1H1613-097, covering an area of 3$^\circ$ in longitude and over 6$^\circ$ in latitude. \hi\ maps of the clouds are shown in Figure \ref{fig:GBT_map}, with the left panel displaying the full field and the right panels providing close-ups of the individual clouds. The clouds exhibit a range of \hi\ column densities spanning one decade ($17.9 \leq \log N_{\rm HI}$/cm$^{2} \leq 18.7$), with a median value of 18.5.

The clouds display complex morphology. Eight of the eleven clouds (A–G and K) are sufficiently large to be resolved by the 9\arcmin\ GBT beam, while the remaining three (H, I, and J) are unresolved. The largest cloud, A, spans approximately 30\arcmin\ in angular size (see Table \ref{:tab:cloud_properties}). The three largest clouds (A–C) exhibit relatively round, symmetric shapes, whereas the smaller clouds (D–G) display elongated structures.

The clouds also show evidence of kinematic substructure, as depicted in Figure \ref{fig:GBT_map_kinematics}, which presents the first-moment GBT maps for each cloud. The corresponding brightness temperature–weighted 1D spectra, along with the fitted Gaussian profiles used to measure the mean \vlsr, are presented in Figure~\ref{fig:GBT_1d_spec}. The larger clouds resolved by the GBT beam (A–E) display a velocity gradient across the cloud, with a variation in \vlsr\ of approximately 10–20 \kms. Cloud A, the largest, has a velocity spread of $\Delta v_{90} \approx$ 18 \kms. In addition, the three elongated clouds (D, E, and G) exhibit significantly larger $\Delta v_{90}$ values of 27, 22, and 19 \kms, respectively. These clouds also appear to be in close proximity to each other in projection (see Figure \ref{fig:GBT_map}, left panel), which may indicate that they are experiencing disturbances likely caused by an external force, such as a nuclear wind. Future observations with higher spatial resolution in 21 cm would be necessary to determine whether these clouds are physically connected and whether they originate from the same parent cloud.

The observed clouds are projected onto the northern Fermi Bubble and exhibit kinematics consistent with the previously identified population of Fermi Bubble high-velocity clouds (HVCs; see Figure~\ref{fig:lsr_profile}). The measured cloud velocities align with theoretical models of outflow from the Galactic nucleus into the Fermi Bubbles. These models describe either a decelerating ballistic wind initially launched at radial velocities of $\approx$ 800–1000~\kms\ \citepalias{Bordoloi2017}, or alternatively, a constant-velocity wind achieving outflow speeds $\geq 400~$\kms\ at distances of approximately 4~kpc from the Galactic center \citep{dT18,Lockman2020}.

All detected \hi\ emission exhibits negative \vlsr\ values. This geometry reflects outflow projection, with negative \vlsr\ arising from the near side of the Fermi Bubbles and positive velocities from the far side (see, e.g., \citealt{Fox2015}). Along the sightline to QSO 1H1613$-$093, the near side lies at $z = 2.5$~kpc and the far side at $z = 6.5$~kpc, based on the model of \citet{Miller2016}. Negative \vlsr\ values are therefore expected, tracing gas closer both to the outflow source and to the Sun. Consistent with this picture, the UV spectra of \citetalias{Bordoloi2017} toward the QSO are also dominated by negative velocities in all species.

The eleven detected clouds show a significant variation in their mean velocities over the angular scale of our survey (approximately 100 \kms\ over a few square degrees; see Table~\ref{:tab:cloud_properties}). This large velocity dispersion over such a small angular extent is atypical when compared to the general HVC population, 
which tend to show smaller velocity variations on these scales \citep{Westmeier2018}.
This observation further supports the interpretation that these \hi\ clouds are related to the nuclear wind rather than the general HVC population. 

Several mechanisms may explain the presence of the observed \hi\ clouds. One possibility is that the \hi\ complex represents debris from a larger, initially coherent cloud that fragmented through interaction with the surrounding hot plasma, potentially via the Kelvin–Helmholtz instability \citep{Scannapieco2015, Scannapieco2017}. Alternatively, the \hi\ clumps may trace neutral gas that has recently condensed out of the hot plasma through thermal instability \citep{Joung2012, Thompson2016}.

Assuming these clouds are entrained gas moving with the nuclear wind, we can compute the cloud survival timescale. The cloud crushing timescale is defined as $t_{\rm cc} = \chi^{1/2} R_{\rm cloud}/v_{\rm wind}$, where $\chi$ is the density contrast between the cold cloud and the surrounding hot wind \citep{Klein1994}. In the absence of strong radiative cooling or magnetic stabilization, cold clouds are typically disrupted within a few $t_{\rm cc}$. However, numerical simulations that include radiative cooling suggest that clouds can survive significantly longer, up to $\sim 10\text{--}20 \, t_{\rm cc}$, before being fully mixed or destroyed \citep{Scannapieco2015}.

Assuming pressure equilibrium between the  $T_{\rm cloud} \sim 10^4 \, \mathrm{K}$ cloud and the hot Fermi Bubble wind with density $n_{\rm hot} \sim 2 \times 10^{-3} \, \mathrm{cm^{-3}}$ and temperature $T_{\rm hot} \sim 3 \times 10^6 \, \mathrm{K}$, the corresponding density contrast is $\chi \sim 300$  \citep{Miller2016}. The highest mass cloud in this work (1H1613\_A) ($M_{\rm cloud} = 1471 \, M_{\odot}$), with  $R_{\rm cloud} = 28 \, \mathrm{pc}$, and wind velocity of the hot wind $v_{\rm wind} \approx 1000 \, \mathrm{km\,s}^{-1}$, we find $t_{\rm cc} \approx 4.3 \times 10^5 \, \mathrm{yr}$, implying a survival time of $\sim 4\text{--}8 \, \mathrm{Myr}$ for this cloud.

For the full cloud population presented in this work, we find a range of cloud crushing timescales with a minimum of $t_{\rm cc}^{\rm min} \approx 4.6 \times 10^4 \, \mathrm{yr}$ ($\sim 0.05 \, \mathrm{Myr}$), a maximum of $t_{\rm cc}^{\rm max} \approx 4.4 \times 10^5 \, \mathrm{yr}$ ($\sim 0.43 \, \mathrm{Myr}$), a mean value of $\langle t_{\rm cc} \rangle \approx 1.8 \times 10^5 \, \mathrm{yr}$ ($\sim 0.18 \, \mathrm{Myr}$). These values imply that typical clouds in the population may survive for $\sim 1\text{--}8 \, \mathrm{Myr}$, assuming disruption occurs on $10\text{--}20 \, t_{\rm cc}$ timescales. This range is consistent with the kinematic age of the Fermi Bubbles derived with direct and indirect methods \citep{Bordoloi2017,Bland-Hawthorn2019,Yang2022,DiTeodoro2020}.

Thermal conduction can significantly reduce cloud survival times, particularly in low-density environments, by promoting evaporation and suppressing radiative cooling. If unsuppressed, classical Spitzer conduction can disrupt clouds on timescales shorter than the cloud-crushing time \( t_{\rm cc} \) \citep{Bruggen2016,Armillotta2017}. These effects place stringent constraints on models that aim to explain the presence of the H\,\textsc{i} cloud complex presented here. The discovery of these high-latitude \hi\ clouds challenges several Fermi Bubble formation models, which require significantly longer formation timescales.

In contrast, magnetic fields can inhibit thermal conduction and suppress instabilities such as Kelvin-Helmholtz and Rayleigh-Taylor, potentially extending cloud lifetimes well beyond $20 \, t_{\rm cc}$ \citep{McCourt2015,Banda-Barragan2016}. Thus, the survival of cold clouds in galactic winds is highly sensitive to the microphysical processes governing wind–cloud interactions, especially the roles of radiative cooling, thermal conduction, and magnetic fields.

The formation and survival of cool clouds in galactic winds remain active areas of theoretical investigation \citep{Scannapieco2017, Schneider2017, Schneider2018, Gronke2018, Farber2022, Drummond2022}. Although a detailed comparison between our Fermi Bubble clouds and these simulations is beyond the scope of this Letter, our observations provide important empirical constraints on the existence and survival of \hi\ clouds at vertical distances $>2$–$4$~kpc from the Galactic center. These results will offer direct tests for the next generation of galactic outflow models.

\begin{deluxetable*}{cccccccccc}
\tablecolumns{10}
\tablewidth{0pt}
\tablecaption{Properties of Individual HVCs detected with the GBT \label{:tab:cloud_properties}}
\tablehead{
\colhead{ID}&
\colhead{$l$}&
\colhead{$b$}&
\colhead{$T_{\rm B,max}$}&
\colhead{$\log N_{\rm HI,max}$} &
\colhead{$v_{\rm LSR}$\tablenotemark{a}}&
\colhead{$\Delta v$\tablenotemark{b}}&
\colhead{$\Delta v_{90}$\tablenotemark{c}}&
\colhead{$M_{\rm HI}$\tablenotemark{d}} &
\colhead{Cloud Size\tablenotemark{d}} \\
\colhead{}&
\colhead{[degr.]} &
\colhead{[degr.]} &
\colhead{[K]}&
\colhead{[$N$ in \cmsq]} &
\colhead{[\kms]}&
\colhead{[\kms]}&
\colhead{[\kms]}&
\colhead{[$M_{\odot}$]} &
\colhead{[pc]}
}
\startdata
1H1613\_A & 3.59 & 28.49 & 0.095 & 18.6 & $-$172  $\pm$ 0.2 & 10 & 18 & 1471 & 27.7 \\ 
1H1613\_B & 4.06 & 30.34 & 0.134 & 18.6 &  $-$96  $\pm$ 0.1 & 12 & 5  &  580 & 18.9 \\
1H1613\_C & 5.12 & 27.24 & 0.138 & 18.6 & $-$124  $\pm$ 0.4 & 7  & 13 &  138 & 12.9 \\
1H1613\_D & 2.66 & 25.81 & 0.217 & 18.6 & $-$157  $\pm$ 0.5 & 9  & 27 &  527 & 19.4 \\
1H1613\_E & 3.21 & 25.84 & 0.096 & 18.5 & $-$174  $\pm$ 0.6 & 12 & 22 &  241 & 15.8 \\
1H1613\_F & 3.69 & 25.53 & 0.066 & 18.3 & $-$176  $\pm$ 0.8 & 10 & 10 &    4.4 &  4.0 \\
1H1613\_G & 2.79 & 26.38 & 0.069 & 18.3 & $-$164  $\pm$ 0.8 & 10 & 19 &   24 &  8.5 \\
1H1613\_H & 4.75 & 27.94 & 0.066 & 18.2 &  $-$95  $\pm$ 0.7 & 7  & 13 &    4.2 &  4.7 \\
1H1613\_I & 3.46 & 27.83 & 0.038 & 17.9 & $-$115  $\pm$ 1.1 & 9  & 6  &    5.0 &  8.5 \\
1H1613\_J & 2.67 & 29.43 & 0.081 & 18.2 & $-$155  $\pm$ 1.3 & 8  & 5  &   11 &  8.8 \\
1H1613\_K & 4.27 & 24.79 & 0.169 & 18.7 &  $-$90  $\pm$ 1.0 & 9  & 4  &   30\tablenotemark{e} &  9.1\tablenotemark{e} \\
\enddata
\tablenotetext{a}{Velocity uncertainty represents the error on the centroid of the Gaussian fit.}
\tablenotetext{b}{Velocity dispersion (1$\sigma$) of the $T_B$-weighted spectrum.}
\tablenotetext{c}{Velocity width of each cloud between the 5th and 95th percentile spaxels.}
\tablenotetext{d}{Assuming circular cloud geometry and a cloud distance from the sun of 7.5 kpc \citepalias{Bordoloi2017}.}
\tablenotetext{e}{Cloud K lies on the edge of our search area, so the \hi\ mass and clouds size are lower limits.}
\vspace{-0.5cm}
\end{deluxetable*}

\section{Summary}\label{sec:summary}

Using deep 21\,cm Green Bank Telescope (GBT) observations of a $3.2^\circ \times 6.2^\circ$ field centered on QSO 1H1613-097 ($l,b = 3.6^\circ,28.5^\circ$), we report the discovery of eleven \hi\ high-velocity clouds (HVCs) in the velocity range $-180$ to $-90$\kms. This survey is more than twice as sensitive as previous \hi\ studies of the Fermi Bubbles, reaching an rms brightness temperature noise of 10mK. Eight of the clouds are spatially resolved, coherent structures with sizes of 4 to 28 pc, peak \hi\ column densities of $\log N_{\rm HI} \sim $ 17.9 to 18.7, and \hi\ masses up to $1470$\msun\ at an adopted distance of 7.5 kpc. The larger clouds exhibit internal velocity gradients and velocity spreads of $\Delta v_{90}\approx$20--30\,\kms. Their central velocities are consistent with UV absorption-line measurements of high-latitude HVCs inside the Fermi Bubbles, and with models of a decelerating nuclear wind launched from the Galactic center \citepalias{Bordoloi2017}.

These clouds represent the highest-latitude 21~cm HVC detections within the Fermi Bubbles to date, extending far beyond the previously studied population near the Galactic center \citep{MG13, dT18, Lockman2020}. Their presence demonstrates that the interior of the Fermi Bubbles is a complex, multiphase environment, with neutral clouds surviving several kiloparsecs above the Galactic plane.

\begin{acknowledgments}
This work is based on data taken in GBT programs 16B-422, 17B-015 and 20A-253. The National Radio Astronomy Observatory is a facility of the National Science Foundation, operated under a cooperative agreement by Associated Universities, Inc. This work is supported by the National Science Foundation under grant number: NSF AST-2206853.  
\end{acknowledgments}

\vspace{5mm}
\facilities{GBT}

\software{astropy \citep{2013A&A...558A..33A,2018AJ....156..123A}, GBTIDL \citep{GBTIDL}, rbcodes \citep{bordoloi_rbcodes_2022}.}

\clearpage
\bibliography{references}{}
\bibliographystyle{aasjournal}

\end{document}